\documentclass[10pt,conference]{IEEEtran}

\usepackage{cite}
\usepackage{url} 

\usepackage [fleqn]{amsmath}
\usepackage{amssymb,amsthm, amsfonts}

\theoremstyle{definition}
\newtheorem{definition}{Definition}[section]

\IEEEoverridecommandlockouts

\usepackage{censor}
\usepackage{microtype}
\usepackage{balance} 

\usepackage{booktabs}   
\usepackage{subcaption} 

\usepackage{etoolbox} 
\usepackage{listings}
 
\usepackage{amsmath,amssymb,mathtools}
\usepackage{color, xcolor}
\usepackage{graphicx}
\usepackage{subcaption}
\usepackage{tabularx} 
\usepackage{amsthm}
\usepackage{wrapfig}

\usepackage{fancybox}

\theoremstyle{plain}

\usepackage{tikz}


\usepackage[nounderscore]{syntax}

\usepackage{mdframed}

\newcommand*{\clz}[1]{\textit{#1}}

\usepackage[ruled]{algorithm2e} 
 
\usepackage{xspace}
\newcommand{\FAR}{FAR\xspace}
\usepackage{stmaryrd}

\newcommand{\domain}[1]{\textit{#1}}

\newcommand{\metaop}[1]{\textit{#1}}

\newcommand{\DEF}{\stackrel{{\rm def}}{=}}

\newcommand{\KW}[1]{\texttt{\textbf{#1}}}
\newcommand{\PREREST}[2]{\KW{pre}~#1; ~\KW{rest}~#2}

\newcommand{\PRE}{\metaop{pre}}
\newcommand{\REST}{\metaop{rest}}

\newcommand{\CASE}{\domain{Case}}
\newcommand{\SPECIFICATION}{\domain{Specification}}
\newcommand{\POWERSET}[1]{{\cal P}(#1)}
\newcommand{\SP}{\metaop{sp}}
\newcommand{\CASES}{\metaop{cases}}
\newcommand{\LETHEAD}[2]{\mbox{{\bf let} }{#1} = {#2}\mbox{ {\bf in} }}

\lstdefinelanguage[JML]{Java}[]{Java}%
       {
        comment=[l]{//\ },
        morecomment=[s]{/*\ }{*/},        
        morecomment=[s]{/**}{*/},
        classoffset=1,
        morekeywords={abrupt_behavior,abrupt_behaviour,
         accessible,accessible_redundantly,also,assert,assert_redundantly,
         assignable,assignable_redundantly,assume,assume_redundantly,
         axiom,behavior,behaviour,breaks,breaks_redundantly,
         callable,callable_redundantly,captures,captures_redundantly,
         choose,choose_if,code,code_bigint_math,code_java_math,
         code_safe_math,constraint,constraint_redundantly,constructor,
         continues,continues_redundantly,decreases,decreases_redundantly,
         decreasing,decreasing_redundantly,diverges,diverges_redundantly,
         duration,duration_redundantly,ensures,ensures_redundantly,
         example,exceptional_behavior,exceptional_behaviour,
         exceptional_example,exsures,exsures_redundantly,extract,field,
         forall,for_example,ghost,helper,hence_by,hence_by_redundantly,
         implies_that,in,in_redundantly,initializer,initially,instance,
         invariant,invariant_redundantly,loop_invariant,
         loop_invariant_redundantly,maintaining,maintaining_redundantly,
         maps,maps_redundantly,measured_by,measured_by_redundantly,method,
         model,model_program,modifiable,modifiable_redundantly,modifies,
         modifies_redundantly,monitored,monitors_for,non_null,
         normal_behavior,normal_behaviour,normal_example,nowarn,
         nullable,nullable_by_default,old,or,post,post_redundantly,
         pre,pre_redundantly,pure,readable,refine,refines,refining,represents,
         represents_redundantly,requires,requires_redundantly,
         returns,returns_redundantly,set,signals,signals_only,
         signals_only_redundantly,signals_redundantly,spec_bigint_math,
         spec_java_math,spec_protected,spec_public,spec_safe_math,
         static_initializer,uninitialized,unreachable,weakly,
         when,when_redundantly,working_space,working_space_redundantly,
         writable
        },
        morekeywords={rep,peer,readonly},
        keywordsprefix=\\,
        otherkeywords={<:,<-,->,..,<==,==>,<==>,<=!=>},
        classoffset=0 
}

\newcommand{\LSTJMLFILESLICE}[2]{
\lstinputlisting[basicstyle=\footnotesize\ttfamily,linerange={#1}]{#2}
}

\tracingpatches
\makeatletter
\patchcmd{\@setref}{\bfseries ??}{\bfseries\color{red} FIXME}{}{}
\makeatother

\newcommand{\cor}{\ensuremath{\lor}}
\newcommand{\cand}{\ensuremath{\land}}

\title{Inferring Concise Specifications of APIs}

\begin{document}

\author{\IEEEauthorblockN{John L. Singleton}
	\IEEEauthorblockA{\textit{accesso Technology Group, plc} \\
		jls@cs.ucf.edu}
	\and
	\IEEEauthorblockN{Gary T. Leavens}
	\IEEEauthorblockA{\textit{University of Central Florida} \\
		leavens@cs.ucf.edu}
	\and
	\IEEEauthorblockN{Hridesh Rajan}
	\IEEEauthorblockA{\textit{Iowa State University} \\
		hridesh@iastate.edu}
	\and
	\IEEEauthorblockN{David R. Cok}
	\IEEEauthorblockA{\textit{GrammaTech, Inc}\\
		dcok@grammatech.com}
}

\maketitle

\thispagestyle{plain} 
\pagestyle{plain}     

\begin{abstract}
Modern software relies on libraries and uses them via
application programming interfaces (APIs).
Correct API usage as well as many software engineering tasks are 
enabled when APIs have formal specifications.
In this work, we analyze the implementation of each method in an API to infer
a formal postcondition.
Conventional wisdom is that, if one has preconditions, then one can use the strongest postcondition predicate transformer (SP) to infer postconditions. 
However, SP yields postconditions that are 
exponentially large, which makes them difficult to use, 
either by humans or by tools.
Our key idea is an algorithm that converts such exponentially large specifications into a form that is more concise and thus more usable.
This is done by leveraging the structure of the specifications that result from the use of SP.  
We applied our technique to infer postconditions for over 2,300 methods in seven popular Java libraries. Our technique was able to infer specifications for 75.7\% of these methods, each of which was verified using an Extended Static Checker. We also found that 84.6\% of resulting specifications were less than 1/4 page (20 lines) in length. Our technique was able to reduce the length of SMT proofs needed for verifying implementations by 76.7\% and reduced prover execution time by 26.7\%. 

\end{abstract}



\section{Introduction}
\label{sec:introduction}

Frameworks and libraries are the basic building blocks of modern software,
using them via their application programming interfaces (APIs),
which are collections of classes and their methods.
A specification for an API method is a contract~\cite{Meyer88,burdy_overview_2005}.
An API method's {\em precondition} is a predicate
that must hold when it is called;
an API method's {\em postcondition} is a predicate that the method ensures will hold when it completes.
For instance, the 
\texttt{push(item)} method of \texttt{java.util.Stack} 
ensures that \texttt{item} is the top of the stack.

Knowledge of postconditions is very useful for automated software 
engineering tools such as formal verification of
program correctness~\cite{Ammons02,ball-spin01,xie-popl05},
test case generation~\cite{dart-pldi05},
test oracle creation~\cite{tonella-fse13}, 
detecting bugs~\cite{engler-sosp01,prminer-fse05,weimer05}, 
design by contract~\cite{burdy_overview_2005,yiwei-icse11}, etc.
Popular formal specification
tools include ESC/Java~\cite{flanagan_extended_2002},
Bandera~\cite{bandera-icse00}, Java Path Finder~\cite{jpf},
JMLC~\cite{leavens_preliminary_2006}, Kiasan~\cite{kiasan-isola06}, Code
Contracts~\cite{codecontracts}, etc., regularly use postconditions
in place of a method call to gain scalability.

Unfortunately, pre- and postconditions specifications 
are not widely available, even for widely-used
libraries. The main reason seems to be that the cost of writing such specifications
is similar to the cost of writing the code itself~\cite{Leavens-Clifton05a}. 
To decrease the cost of writing specifications, several sets of techniques have
been proposed to automatically derive specifications. 

A first set of techniques analyzes call sites of an API method
to collect a set of predicates at each of these call sites and then uses mining 
techniques, such as frequent items mining, to infer preconditions
\cite{nguyen_mining_2014-1,Ramanathan07}.
Another body of work has focused on analyzing call sites to mine temporal patterns
over API method calls, e.g. \cite{Gruska-Wasylkowski-Zeller-10,Nguyen09,Wasylkowski07,Livshits05,Williams05,Michail00}.
However, these works do not infer postconditions.

A second set of techniques uses static analysis on the code 
of the API method to infer specifications; e.g. Cousot {\em et al.} uses 
abstract interpretation~\cite{Cousot2013} to infer preconditions, 
and Buse {\em et al.} uses symbolic execution~\cite{Buse2008} to 
infer conditions leading to exceptions.
However, Buse {\em et al.} do not infer conditions under which the API method 
terminates normally and Cousot {\em et al.} do not infer postconditions.

A third set of techniques uses {\em dynamic} approaches to mining
specifications~\cite{Ammons02,Cousot2011,Dallmeier05,ernst_daikon_2007,Gabel08,Liu06,Lo09,Pradel09,Wei11,Yang06}.
Some of these works infer temporal patterns over API method calls~\cite{Yang06,Liu06,Gabel08,Lo09},  
object-usage specifications~\cite{Pradel09},
and others 
strengthen existing specifications~\cite{Wei11}.
Although Daikon~\cite{ernst_daikon_2007} can infer postconditions, 
its inference depends on the presence of an adequate test suite~\cite{harder_improving_2003,nimmer_invariant_2002}. 

\begin{figure*}[tp] 
	   
	\begin{subfigure}[t]{0.33\linewidth}
		\LSTJMLFILESLICE{10-26}{listings/cmp.java}
		\caption{A Java function \texttt{cmp}}  
		\label{fig:cmp}
	\end{subfigure}
	\begin{subfigure}[t]{.335\linewidth}
		\LSTJMLFILESLICE{2-23}{listings/cmp-contract-pre.java}
		\caption{Postcondition of \texttt{cmp} using the standard SP} 
		\label{fig:cmp-pre}
	\end{subfigure}
	\hspace*{\fill}
	\begin{subfigure}[t]{.33\linewidth}
		\LSTJMLFILESLICE{5-26}{listings/cmp-contract-post.java}
		\caption{Postcondition using our technique}
		\label{fig:cmp-post} 
	\end{subfigure}
	\caption{The postcondition of \texttt{cmp} function obtained 
	using the standard definition of SP is verbose, exposes implementation details,
	and is tied to the structure and flow of \texttt{cmp}.
	Our technique infers concise postconditions that eliminates
	those limitations.
}
    \label{fig:cmp-overall}
	
\end{figure*}

This paper proposes a technique for inferring postconditions that combines
forward symbolic execution~\cite{gordon_forward_2010} with predicate transformers~\cite{dijkstra_discipline_1997}.
Starting from a precondition (e.g., \lstinline!true! or one inferred 
using prior techniques~\cite{nguyen_mining_2014-1,Ramanathan07}) 
our technique uses the body of the API method to produce a logical 
formula that can be converted into a specification, as shown in Figure~\ref{fig:cmp-overall}.

To empirically evaluate our approach, we apply it to infer postconditions 
for seven popular Java libraries: JUnit4 (JU4), JSON-Java (JJA), Commons-CSV (CSV), Commons-CLI (CLI), Commons-Codec (COD), Commons-Email (EMA), and Commons-IO (CIO) 
totaling over 2300 methods.
Our results show that our technique has a very high precision and recall.
We were able to infer specifications for 75.7\% methods, and 
all of the inferred specifications were verified to be correct using
OpenJML's Extended Static Checker (with Z3 \cite{de2008z3} version 4.3.0).
Our results show that our inferred specifications are both rich and concise.
To evaluate richness, we study the presence of \texttt{assignable} and 
\texttt{purity} clauses (in addition to pre- and postconditions) in inferred 
specifications. The purity clause documents whether an API method will change
memory locations, and the assignable clause documents which memory locations
can be changed. 
We find that all the API methods with inferred specifications have either 
a purity clause or an assignable clause. 
To evaluate conciseness, we study the length of final specifications and find 
that 84.6\% of inferred specifications are less than 1/4 page in length. 

Our key contributions include: 
\begin{enumerate}
	\item A novel technique for mitigating the effect of state space explosion in the context of inferring specifications using strongest postcondition predicate transformers.
	\item The implementation of a tool, {\bf Strongarm}, that
	embodies our postcondition inference technique as an extension of the OpenJML \cite{leavens_preliminary_2006,burdy_overview_2005} program
	verification tool. Our tool will be part of a future OpenJML release
	that will make our techniques available to formal methods community.
	\item A series of techniques and a novel, graph-based algorithm 
	for producing concise specifications.
	\item An experimental evaluation of Strongarm on seven
	popular Java libraries that demonstrates its effectiveness.
	\item A benchmark of seven specified libraries.
\end{enumerate}

Next, we will motivate our approach. 
Section~\ref{sec:our-approach} describes our approach, 
Section \ref{sec:technical-evaluation} presents its evaluation,
Section \ref{sec:related-work} compares it with related work, 
and Section \ref{sec:conclusion} concludes.

\section{Motivation}
\label{sec:motivating-example}

The standard technique for statically computing the postcondition for a piece of 
code was formalized by Dijkstra in the form of the
\textit{strongest postcondition predicate transformer} \cite{dijkstra_discipline_1997}
(SP). SP is a functions that takes a predicate describing a starting
state (i.e., a pre-state) and a statement, and produces predicate that describes
the set of states that may result from executing that statement on any
pre-state that satisfies the given predicate.
Many researchers have used the opposite predicate transformer, the
weakest precondition transformer, also introduced by Dijkstra, 
for building verification tools,
including ESC/Java, Boogie \cite{leino_this_2008}, OpenJML, and many others. 
However, SP itself has not been previously leveraged for successful tools that infer
postconditions.

To understand reasons why SP has not been successfully used before, 
consider inferring the postcondition of the Java function in Figure \ref{fig:cmp}.
Figure \ref{fig:cmp-pre} shows the results of applying SP on that
code, starting from the predicate \textbf{\texttt{true}},
and translating the resulting proposition to the JML notation.
The JML keyword normal\_behavior means that this specification is
about the normal termination.
As is usual in SP, conditions are accumulated as the tool traverses the control flow
of the function.
Since this function has three exits, this results in three alternative postconditions.
The syntax \{$\mid$~A~\texttt{\textbf{also}}~B~$\mid$\} means that the
implementation must obey both specifications $A$ and $B$,
or, equivalently, that a caller may conclude the postcondition of either $A$ or $B$.
Starting with the precondition of \textbf{\texttt{true}}, 
the first assignment leads to accumulation of the condition \texttt{c==a}.
The if-condition leads to accumulation of the condition \texttt{c<b} in the 
true branch, and its negation \texttt{!(c<b)} in the false branch.
The first (lexical-order) return statement is an exit point, and 
thus leads the postcondition \texttt{\textbackslash result} \texttt{==} \texttt{-1}.
Thus, at the first return statement we get four accumulated conditions. 
\texttt{c<b} (from the if condition), \texttt{true} (from the precondition), 
\texttt{\textbackslash result} \texttt{==} \texttt{-1} (from the
return), and \texttt{c==a} (from the assignment).
The second if statement (within the body of else) leads to accumulation of the 
condition \texttt{c>b} in the true branch and its negation in the false branch.
The second return statement thus
leads to the postcondition \texttt{\textbackslash result} \texttt{==} \texttt{1},
along with 
\texttt{!(c<b)} (from the first if), \texttt{(c>b)} (from the second if), 
and \texttt{true} (from the precondition).
The third return statement is similar.

As the resulting postconditions show,  
a straightforward use of the predicates produced by SP for
specification inference
leads to verbose results that: (1) expose the internal details of the function
such as internal variables (e.g., \texttt{c}) and assignments to such variables that are 
not part of the interface, 
(2) postconditions that are directly tied to the structure of the code
and its control flow, and
(3) significant redundancies that could increase the cost of reading and using
such specifications.

Figure \ref{fig:cmp-post} shows the postconditions produced by our approach
and previews its advantages.
These postconditions are in terms of function arguments, 
reorganized to reduce redundancies, and eliminate internal 
details and tautologies.




\begin{figure*}[t] 
 
\centering 
\includegraphics[width=\linewidth]{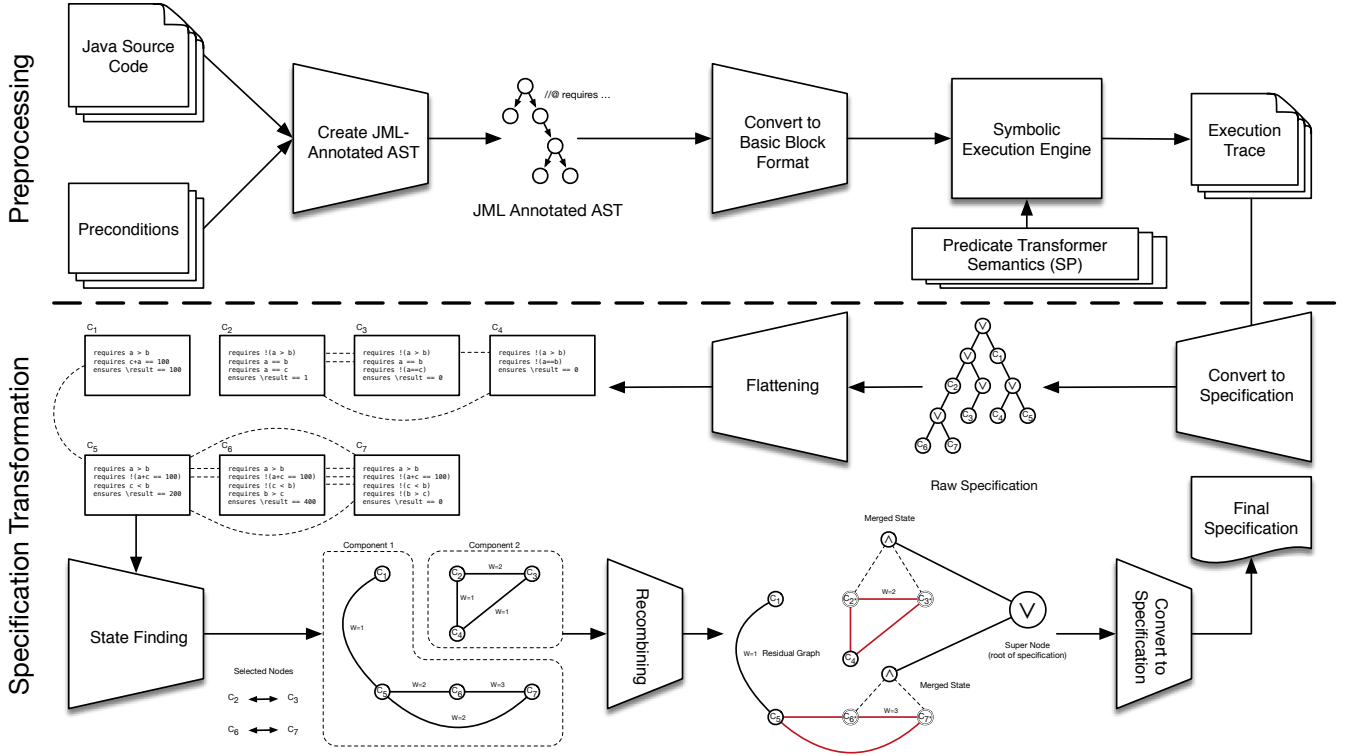}

\caption{Approach overview: The top-half produces raw specifications, and the lower half makes them concise.}
 
%
\label{fig:overview-pipeline}

\end{figure*}



\section{Inferring Concise Postconditions}
\label{sec:technical-approach}
\label{sec:our-approach}

Figure \ref{fig:overview-pipeline} shows an overview of our approach for 
inferring postconditions that consists of the following steps.
\begin{enumerate}
	\item The input is the code of the API method for which postconditions are needed. 
	      If preconditions are available, they can also be provided. 
	      Otherwise, our approach starts with the default precondition \textbf{\texttt{true}}.
	\item Next, the code is symbolically 
	      executed to produce traces. The symbolic execution uses the strongest postcondition predicate transformer semantics (SP). 
	      These traces are converted to a raw specification. (See Sec.~\ref{subsec:sp} below.) 
	\item Then, we flatten the raw specification into groups of clauses (cases). 
		  (See Sec.~\ref{subsec:flattening})
	\item Next, we compute the overlapping states found between groups of clauses. (See Sec.~\ref{subsec:state-finding}.)
	\item Finally, we recombine states and convert the results to
          the final specification. (See Sec.~\ref{subsec:recombining}.)
\end{enumerate}

\subsection{Producing Raw Specifications}
\label{subsec:sp}

We use forwards symbolic execution \cite{gordon_forward_2010} to symbolically
execute the annotated AST of each API method. 
The rules used by our symbolic execution engine are shown in Fig.~\ref{fig:rules}.

\begin{figure}[t]
  \setlength{\mathindent}{0pt} 
    \begin{align}
      & \texttt{sp SKIP}~ P  =  P  \\
      & \texttt{sp}~ (V := E )~ P = \exists v.(V=E[v/V]) \cand P[v/V] \label{eqn:ass} \\
      & \texttt{sp}~ (S_1; S_2 )~ P = \texttt{sp}~ S_2 (\texttt{sp}~ S_1 P) \\
      & \nonumber \texttt{sp}~ (\texttt{IF}~ B~ \texttt{THEN}~ S_1~ \texttt{ELSE}~ S_2)~ P =  (\texttt{sp}~ S_1 (P \cand B))  \cor \\ &\quad\quad\quad\quad\quad\quad\quad\quad\quad\quad\quad\quad\quad (\texttt{sp}~ S_2 (P \cand \lnot B)) \label{eqn:if} \\
      & \nonumber \texttt{sp } (\texttt{WHILE } B \texttt{ DO } S)~ P =  (\texttt{sp } (\texttt{WHILE } B \texttt{ DO } S)  \\
       & \quad\quad\quad\quad\quad\quad\quad\quad\quad\quad(\texttt{sp } S~ (P \cand B)))~\cor (P \cand \lnot B)
    \end{align}

%


\caption{Predicate transformer rules used by our approach. }
\label{fig:rules}
\end{figure}



A challenge in this step was to avoid existential quantifiers in SP's output.
Existential quantifiers are problematic because they require the use of a constraint solver.
However, the assignment rule in Equation \ref{eqn:ass} in Fig.~\ref{fig:rules} uses an existential quantifier.  We avoid this problem by using a type of program representation called the ``Optimal Passive Form'' \cite{grigore_strongest_2009,leino_this_2008,barnett_weakest-precondition_2005}. 
In Optimal Passive Form, which is a variant of single static assignment (SSA) form,
in which every assignment to a variable results in a new variable.
In addition to simplifying the control flow graph of a program, this form eliminates the need for such existential quantifiers since one does not need to search for previous assignments of variables when making new assignments. 
Next, we convert the propositions produced by SP into a raw specification.
We define an abstract form of specifications independent of any concrete specification language, 
such as JML. 
The specifications themselves are to be thought of as method specifications.
The only interesting properties we assume about specifications are that one can extract cases (such as those produced by $SP$) from a specification
and that each such case has a set of preconditions (conceptually a conjunction) and a set of other clauses (also conjoined).
Thus each specification is modeled as a non-empty list of cases, separated by a special operator $\vee$.  
Within a case, clauses such as preconditions and postconditions are modeled as ``atomic formula,''
assertions whose structure is not further examined.
This step produces specifications in a form that we call {\em specification normal form} (SNF).
While specifications allows preconditions to distribute over cases, 
in SNF no such distribution of preconditions is allowed;
thus in SNF each case is a non-empty list of cases, each separated by the operator $\vee$. 

\begin{definition}[Specification Normal Form (SNF)]
	\label{def:snf}
	A specification is in \emph{specification normal form (SNF)} if it follows the grammar in Figure~\ref{fig:snf-grammar}.
	\setlength{\grammarparsep}{2pt} 
\setlength{\grammarindent}{6em} 

\begin{figure}[ht]
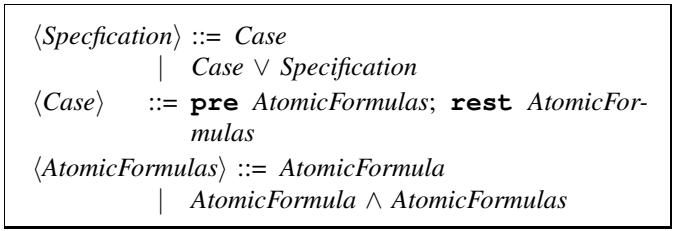


  \begin{mdframed}
    \begin{grammar}

    <Specfication> ::= \clz{Case} \alt \clz{Case} $\vee$ \clz{Specification}

    <Case> ::= \KW{pre} \clz{AtomicFormulas}; \KW{rest} \clz{AtomicFormulas}

    <AtomicFormulas> ::= \clz{AtomicFormula} \alt \clz{AtomicFormula} $\wedge$ \clz{AtomicFormulas}
    \end{grammar}
  \end{mdframed}
  
  \caption{The abstract syntax of specification normal form.}
  \label{fig:snf-grammar}
\end{figure}

\end{definition}


To illustrate this phase of our technique consider the function in Figure~\ref{fig:cmp}.
The computation of $\SP(\texttt{cmp},\phi)$ produces the following formula:

\begin{equation}\label{eqn:sp-computed}
\begin{array}{l}
(a < b \land \phi \land \KW{result} = -1) \\
~ \lor \: ( (a > b \land \lnot(a < b) \land \phi \land \KW{result} = 1) \\
~ \lor \: (\lnot(a < b) \land \lnot (a > b) \land \phi \land \KW{result} = 0))
\end{array}
\end{equation}

Formula~(\ref{eqn:sp-computed}) corresponds to the following specification.

\begin{equation}\label{eqn:sp-computed-spec}
\begin{array}{l}
(\PREREST{a < b \land \phi}{\KW{result} = -1}) ~\vee \\
\: (\PREREST{a > b \land \lnot(a < b) \land \phi}{\KW{result} = 1}) ~\vee \\
\: (\PREREST{\lnot(a < b) \land \lnot (a > b) \land \phi}{\KW{result} = 0})
\end{array}
\end{equation}

\subsection{Flattening Specification Cases and Computing Weights}
\label{subsec:flattening}

\begin{algorithm}[t]
    \LinesNumberedHidden  
    \DontPrintSemicolon
    \SetKwFunction{Cases}{Cases} 
    \SetKwFunction{ToGraph}{ToGraph}
    \SetKwFunction{StronglyConnectedComponent}{StronglyConnected}
    \SetKwFunction{MergeSCC}{MergeSCC}
    \SetKwFunction{ToSpecification}{ToSpec}
    \SetKwFunction{EmptyGraph}{EmptyGraph}
    
    \SetKwData{S}{S}
    \SetKwData{Sim}{$\sim$}

    \KwIn{$\S$, a specification in SNF \newline 
    $\Sim$, a sound equivalence relation
    }
    \KwOut{A specification}
    \Begin{
      \ShowLn $V \leftarrow \CASES(\S)$ \label{ln:far-inputspecification}\;
      \ShowLn $G \leftarrow \EmptyGraph{}$ \label{ln:inv-established}\;

       \emph{// merge the overlapping state spaces of \S}
      \;
      \ShowLn \label{ln:far-loop} \Repeat{$|G''| = 0$}{
        \ShowLn $(G',W) \leftarrow \ToGraph{V}$ \emph{// (Alg. \ref{alg:to-graph})} \label{ln:far-loop1}\;  
        \ShowLn $G'' \leftarrow \StronglyConnectedComponent{G'}$ \label{ln:far-loop2}\;
        \ShowLn $(G,V) \leftarrow \MergeSCC{V,W,$G''$,G,\Sim}$ \emph{// (Alg. \ref{alg:mergescc})} \label{ln:far-loop3}\;
      }
      \emph{// connect each unmerged vertex to the root}
      \;
      \ShowLn $G.E = G.E \cup \{(G.root, v) ~|~ v \in V\}$ \label{ln:far-makeg-1}\;
      \ShowLn $G.V = G.V \cup \{V\}$ \label{ln:far-makeg-2}\;
      
     \emph{// convert the residual graph back to a specification} \;
     \ShowLn \Return{$\ToSpecification{G, G.root}$} \emph{// (Alg. \ref{alg:to-specification})}\;    
      }
    
\caption{FAR\label{alg:far}}
\end{algorithm}

\begin{algorithm}
  \LinesNumberedHidden  
  \DontPrintSemicolon
  \SetKwFunction{Cases}{Cases}
  \SetKwFunction{ToGraph}{ToGraph}
  \SetKwFunction{StronglyConnectedComponent}{StronglyConnected}
  \SetKwFunction{MergeSCC}{MergeSCC}
  \SetKwFunction{ToSpecification}{ToSpec}
  \SetKwFunction{EmptyGraph}{EmptyGraph}
  
  \SetKwData{V}{V}
  \SetKwData{Sim}{$\sim$}

  \KwIn{\V, a set of specification cases \newline 
  $\Sim$, a sound equivalence relation
  }
  \KwOut{A graph $G$ and a table of weights $W$}
  \Begin{
    \ShowLn $G \longleftarrow \EmptyGraph()$ \;

    \ShowLn \For{$(l,r) \in \V \times \V $ such that $l \neq r$}{
              \emph{// add $l$ and $r$ to the list of vertices} \;      
    \ShowLn   $G.V \longleftarrow G.V \cup \{l\} \cup \{r\}$ \;
     
              \emph{// compute the weight modulo \Sim} \;
    \ShowLn   $W[l][r] \longleftarrow |\{ (x,y) \in \PRE(l) \times \PRE(r) ~|~ x \Sim y\}|$ \label{ln:weights} \;
 
             \emph{// $l \rightarrow r $ if they share any clauses in common} \;

    \ShowLn   \label{ln:to-graph-branch} \If{$W[l][r] > 0$}{
    \ShowLn     $G.E \longleftarrow G.E \cup \{(l, r)\}$ \;
     }

    }
    \ShowLn \Return{$(G,W)$}
  }

  \caption{ToGraph\label{alg:to-graph}}
\end{algorithm}

\begin{algorithm}
  \LinesNumberedHidden
  \DontPrintSemicolon
  \SetKwFunction{GetCases}{GetCases}
  \SetKwFunction{ToGraph}{ToGraph}
  \SetKwFunction{StronglyConnectedComponent}{StronglyConnectedComponent}
  \SetKwFunction{MergeSCC}{MergeSCC}
  \SetKwFunction{ToSpecification}{ToSpec}
  \SetKwFunction{EmptyGraph}{EmptyGraph}
  
  \SetKwData{V}{V}
  \SetKwData{W}{W}
  \SetKwData{Scc}{Scc}
  \SetKwData{Residual}{R}
  \SetKwData{Sim}{$\sim$}
  
  \KwIn{\V, a set of specification cases  \newline \W, a table of weights between edge pairs of \Scc \newline \Scc, the strongly connected component, a set of graphs of the form $\langle V,E \rangle$ \newline \Residual, a residual graph of the form $\langle V,E \rangle$ \newline \Sim, a sound equivalence relation
  }

  \KwOut{A pair consisting of the \Residual and remaining vertices \V}

  \Begin{
   \ShowLn \label{ln:merge-loop} \For{$G \in Scc$ such that $|G.V| > 1 $}{
   \ShowLn   $maxW \longleftarrow $ the max weight of $G$ via \W \label{ln:getMax} \;
   \ShowLn   $(L,R) \longleftarrow $ a pair $\in G.E$ with weight $maxW$ \label{ln:select-pair} \;
             \emph{// compute intersection modulo \Sim} \;      
   \ShowLn   $c \longleftarrow \{ l ~|~ (l,r) \in \PRE(L) \times \PRE(R)$,  $l \Sim r\} $ \label{ln:common} \;
   \ShowLn   $rcmn \longleftarrow \{ r ~|~ (l,r) \in \PRE(L) \times \PRE(R)$,   $l \Sim r\} $ \label{ln:rcommon} \;

             \emph{// remove the common vertexes} \;
   \ShowLn   $L' \longleftarrow \PREREST{(pre(L) \setminus c)}{rest(L)}$ \label{ln:Lprime} \;
   \ShowLn   $R' \longleftarrow \PREREST{(\PRE(R) \setminus rcmn)}{\REST(R)} $ \label{ln:Rprime} \;
 
             \emph{// make a conjunction node between $L$ and $R$} \;
   \ShowLn 
   \ShowLn  \label{ln:replace-v1} $\Residual.E \longleftarrow \Residual.E \cup (\{c\} \times \{L', R'\}) \cup \{(\Residual.root, c)\}$\;
   \ShowLn   \label{ln:replace-v2} $\Residual.V \longleftarrow \Residual.V \cup \{L'\} \cup \{R'\}$ \label{ln:mergescc-replaceV} \;

             \emph{// remove merged nodes} \;
   \ShowLn   $\V \longleftarrow \V \setminus (\{L\} \cup \{R\})$ \label{ln:mergescc-remove-v}\;
   }

   \ShowLn \Return{$(\Residual,\V)$}
  }

\caption{MergeSCC\label{alg:mergescc}}
\end{algorithm}

\begin{algorithm}[t]
    \LinesNumberedHidden  
    \DontPrintSemicolon
    
    \SetKwData{Residual}{R}
    \SetKwData{Root}{Root}
    
    \KwIn{\Residual, a residual graph of the form $\langle V,E \rangle$ \newline \Root, the root of the graph}

    \KwOut{A specification}

    \Begin{
      \ShowLn $B \longleftarrow \emptyset$ \;
      \ShowLn \label{ln:foreach-adjacent} \For{each $v$ adjacent to \Root}{
      \ShowLn \label{ln:isaleaf} \uIf{$v$ is a leaf}{
        \ShowLn $B \longleftarrow B \cup v$ \;
      }\Else{
        \ShowLn $B \longleftarrow B \cup (v \land \ToSpecification(\Residual, v))$ \;
      }
      
      }
  
      \ShowLn \Return{$\bigvee B$} \label{ln:return-tospecification} \;
    }
  \caption{ToSpec \label{alg:to-specification}}
  \end{algorithm}







Algorithm \ref{alg:far} describes the next three components of our technique.
The input of this component is the raw specification produced by the predicate transformer semantics in SNF. 

In the first step of the algorithm,
these clauses are placed in a data structure suitable for a graph-based analysis; 
our implementation uses the JPaul program analysis library \cite{jpaul}.

First we collect the cases of the specification, represented as $\CASES$ in 
Algorithm \ref{alg:far}, as a set. 
The semantics of case extraction is as follows.
\begin{displaymath}
\begin{array}{l}
\CASES : \SPECIFICATION \rightarrow \POWERSET{\CASE} \\
\CASES(S) \DEF \LETHEAD{c_1 \vee \ldots \vee c_n}{S} \{ c_1, \ldots, c_n \}      
\end{array}
\end{displaymath}

With the cases of the specification collected, the next step is to flatten the specification. This is done by taking each case of the raw specification and detaching it from the tree. This forms a disjoint forest of all the specifications cases. 

Once flattening is complete, Strongarm computes weights for each specification case
using an equivalence relation on pairs of distinct cases as shown in Algorithm \ref{alg:to-graph}.
On Line 7 of Algorithm \ref{alg:to-graph} this equivalence relation is denoted by the symbol `$\sim$'. For simplicity, our prototype uses lexical identity for the relation $\sim$. 
In our experimental analysis (Section \ref{sec:technical-evaluation}) we find that this choice works out well.\footnote{The $\sim$ relation used need not be lexical identity; we speculate that more interesting results could be achieved by considering other relations, such as logical equivalence, but that is beyond the scope of this paper.}
The weights assigned are the number of preconditions that are $\sim$ between the two cases;
for example, if cases $l$ and $r$ have 3 preconditions that are related by $\sim$, then the weight assigned to the edge from $l$ to $r$ is 3.
In Fig.~\ref{fig:overview-pipeline}, an edge of weight $n$ is shown as $n$ dashed edges.

\subsection{State Finding}
\label{subsec:state-finding}

Returning to Algorithm \ref{alg:far},
the next step is to compute the connected components of the resulting graph. In Figure \ref{fig:overview-pipeline}, this process produces Component 1 and Component 2.

\subsection{Recombining and Conversion to Contract}
\label{subsec:recombining}

The next step of Algorithm \ref{alg:far} selects, from each connected component, the pair of vertices with the largest weight edges connecting them (i.e., the most similarity), by calling \texttt{MergeSCC} (Algorithm \ref{alg:mergescc}).
For example, in Figure \ref{fig:overview-pipeline} this produces the choice $(C_6,C_7)$ for Component 1 and $(C_2,C3)$ for Component 2. 

Algorithm \ref{alg:mergescc} works on each connected component as follows.
First it selects from the connected component, the maximum weight (Line~3), 
and a pair of vertices connected by an edge with that weight (Line~4).
The selected pair of verticies are combined into a single node by extracting the common preconditions between the two verticies
(Line~6) and returning a modified residual graph with an edge connecting the common preconditions
($c$) to both verticies without those common specifications ($L'$ and $R'$, Lines~9 to 13).
For example, in Fig.~\ref{fig:overview-pipeline} $C_2$ and $C_3$ are combined into a single node, as are $C_6$ and $C_7$. 
Next, we examine the connected component to which the pair belongs. If any vertex is adjacent to any of the elements of the pair, we remove that edge from the graph (Line~15).

Upon returning to Algorithm \ref{alg:far}, Lines~11 to 12 make a root node for the specification.
When converted to a specification, this root node will be a disjunction ($\vee$) of all the newly-created conjunction nodes 
(the merged states in Fig.~\ref{fig:overview-pipeline}). In the example in Figure \ref{fig:overview-pipeline}, this leaves us with vertices $C_1$ and  $C_5$.
This subgraph is then fed back into the FAR algorithm until there are no remaining vertices. Completely disjoint vertices (those with no preconditions in common with any of the other vertices) form a disjoint forest in the final iteration of the algorithm. Once this outcome is reached, each vertex in the forest is connected unmodified to the root. This resulting tree is then converted back into a specification. 

\subsection{Soundness of Our Approach}
\label{subsec:soundness-brief}

We have proved the soundness of our approach in a detailed report.
Briefly, soundness for our approach is defined with respect to satisfaction by programs (method bodies). 
In essence a program $C$ satisfies a specification $S$ if for every pre-state $s$ that satisfies some case $c$'s precondition, the semantics of $C$ is such that the post-state $s'$ that $C$ produces (when run on $s$) satisfies $c$'s postcondition.
The $\FAR$ algorithm uses an equivalence relation $\sim$ to compute
the intersection between two sets of preconditions. 
However, not just any equivalence relation on atomic formulas will do; a \emph{sound} relation must preserve the meaning of atomic formulas.
Lexical equivalence is sound. Our report shows that if $\sim$ is sound, then
the semantics of specifications is preserved by Algorithm \ref{alg:far}.

\subsection{Implementation Details}
\label{subsec:implementation-details}
After an execution trace of an API method is obtained, we need to transform the trace and the underlying predicate AST into a raw specification. 
Our implementation includes several transformation tasks such as translation from internal variable representations to externalized (source code level) representations, removal of tautologies, determination of purity, and inference of frame axioms. For space considerations we omit a detailed description of these techniques. 

\section{Technical Evaluation}
\label{sec:technical-evaluation}

We have implemented our techniques in a tool, {\bf Strongarm}, that
is an extension of the OpenJML \cite{leavens_preliminary_2006,burdy_overview_2005} 
program verification tool. Our tool will be part of a future OpenJML release.
In this section we evaluate the performance of our technique on the task of inferring specifications. In the next subsection we explain our experimental setup and provide specific details about the source code we conducted our experiments on. Following this, our evaluation looks at the performance of our technique from the following five perspectives:

\begin{enumerate}

  \item \textbf{Effectiveness of Inference} How many of the candidate methods we were able to infer?

  \item \textbf{Efficiency of Reduction} By how much were the specifications reduced?

  \item \textbf{Complexity of Inferred Specifications} Overly complex and overly simplistic specifications are not practical. What are the characteristics of the inferred specifications?

  \item \textbf{Performance of Inference Procedure} How well (with respect to time) did our technique perform at the task of inferring specifications?

  \item \textbf{Performance of Inferred Specifications} How effective is our technique at reducing specification nesting, decreasing specification length, decreasing prover execution time, and decreasing proof length?
    
\end{enumerate}

Finally, in Section \ref{ln:practical} we conclude with a human evaluation of the inferred specifications produced by our technique. 

\subsection{Evaluation Methodology}
\label{sec:data-collection}

We designed a series of experiments designed to examine the effectiveness of our technique at inferring practical specifications. We selected a cross section of popular Java libraries: JUnit4 (JU4), JSON-Java (JJA), Commons-CSV (CSV), Commons-CLI (CLI), Commons-Codec (COD), Commons-Email (EMA), and Commons-IO (CIO). 
The static characteristics of these libraries are summarized in Table~\ref{table:code}. 

\renewcommand{\arraystretch}{1.1}
\begin{table}[h]
	\caption{Code metrics for APIs used in evaluation.}
	\label{table:code}
  \begin{tabularx}{\linewidth}{Xrrrr}
    \toprule  
    API Name &  SLOC & Methods & Files & Version  \\ \toprule
    JUnit4 & 10,018 & 1,230 & 193 & 4.13\\
    JSON-Java & 3,201 & 200 & 18 & 20160212 \\
    Commons-CSV & 1,501 & 158 & 10 & 1.4 \\
    Commons-CLI & 2,666 & 194 & 22 & 1.3.1 \\
    Commons-Codec & 6,607 & 509 & 60 & 1.10\\
    Commons-Email & 2,734 & 192 & 22 & 1.4\\
    Commons-IO & 9,836 & 955 & 115 & 2.5 \\
    \midrule 
    Total & \textbf{36,563} & \textbf{2,331} & \textbf{440} & -  \\
    \bottomrule 
  \end{tabularx}
\end{table}

A method for inferring preconditions in large corpora was recently investigated in the work of Nguyen  et al. \cite{nguyen_mining_2014-1}. However, rather than specifying the preconditions for the methods in our study, we assume a vacuously true precondition, namely \lstinline!true!. Prior to construction of the final specification, the default precondition is removed and not tabulated in the final specification analysis. We do this so as to not simultaneously test the results of our work in parallel with the technique of Nguyen. As noted in Section \ref{sec:limitations}, we do not specifically attempt to infer invariants for loops and instead rely on user-written loop invariants. For the study presented in this paper loops were not annotated with such invariants. Additionally, our technique's inference technique assumes any field referenced in the body of a method should be visible in the inferred specification. However, in JML this is considered an error by default. For this reason, all private fields referenced in specifications have been given the special annotation \texttt{spec\_public} which allows private fields to appear in specifications. This promotion of private fields to \texttt{spec\_public} is performed automatically by our technique during inference and reflected in the inferred specifications; future work includes using JML features such as model fields to avoid declaring all fields to be \texttt{spec\_public}. Additionally, during our evaluation we discovered there were several methods we were unable to validate due to bugs in OpenJML; these methods were explicitly skipped in our evaluation. 
  
Experiments were performed on the Stokes HPC cluster\footnote{https://arcc.ist.ucf.edu/index.php/resources/stokes/about-stokes} at the University of Central Florida. Each job node was configured with 6 Intel Xeon 64-bit processors, 42GB of RAM, and used Oracle JDK 1.8.0.131. Our technique produces extensive telemetry data as well as the inferred specifications. The results presented below are based on mining this telemetry data.

\subsubsection{Verification of Inferred Specifications}
\label{sec:ground-truth-constr}


Once a specification is inferred, we must have a standard way of knowing if the specification itself is correct. In our experiments we validated the inferred specifications in three different ways. First, in the creation of our technique, we built a comprehensive test suite consisting of approximately 100 hand-written and hand-verified test cases. Our tool passes this test suite. Second, once inferred specifications are produced, we use OpenJML's to type check the produced specifications. All of the specifications produced by our technique type check. Lastly, we use OpenJML's Extended Static Checker (with Z3 \cite{de2008z3} version 4.3.0).
70\% of the inferred specifications were verified by OpenJML.
In practice our technique would check these specifications before submitting them to a user and therefore would not produce an invalid specification. The cases we were unable to verify were caused by tool error or implementation issues in our technique. Verification in this fashion also checks for unsatisfiable clauses, which should not be present in practical specifications. In doing this, as noted in Section \ref{sec:limitations}, we do not consider the exceptional behavior of the method in question.  

\subsubsection{Threats to Validity} 
\label{sec:threats-validity}
As mentioned in Section \ref{sec:ground-truth-constr}, we use the OpenJML tool to check the results of our inferred specifications. A positive result from OpenJML certifies that the program code satisfies the given specification. This is an especially strong guarantee; since checking is done statically, this certifies that \textit{for all runs} (and potential input values) the specifications are valid. However, this depends on the soundness of the tool. This is a limitation in the following ways. First, OpenJML itself might have bugs and therefore might certify programs that are not correct. Second, the theory behind OpenJML itself might not be sound, i.e., it might admit incorrect programs (programs that do not satisfy their specification) under certain circumstances; however, it is believed that this second problem is limited to JML features with semantics that are not quite settled yet (such as details concerning invariants). Our study is not impacted by known problems in the semantics of JML. Lastly, in choosing JML as our target specification language, our approach to specification inference may not generalize to other specification languages as well as it has with JML. However, there are many Hoare-style specification languages that use pre- and postcondition specifications, such as Eiffel, and retargeting our technique to these specification languages should be a straightforward exercise for future work.

\subsection{Limitations}
\label{sec:limitations}
In designing our technique, we intentionally made some trade-offs to simplify its implementation. The two major limitations of our technique are seen in our handling of exceptions and in our handling of loop constructs. First, although JML allows for descriptions of exceptional method behavior, in our technique we do not attempt to infer this behavior (although information about the exceptional behavior of codes is present in our AST; it is simply elided). Instead this task is relegated to future work \cite{Dillig2013}. Second, as a simplifying assumption, in our implementation we assume that in the presence of loops that our technique will always have access to expert-written loop invariants. From these loop invariants we apply the standard Hoare loop rules to facilitate inferring postconditions.

\subsection{Effectiveness of Inference}
\label{sec:effect-infer}

\begin{figure}[t]
  \centering
  \includegraphics[width=.8\linewidth]{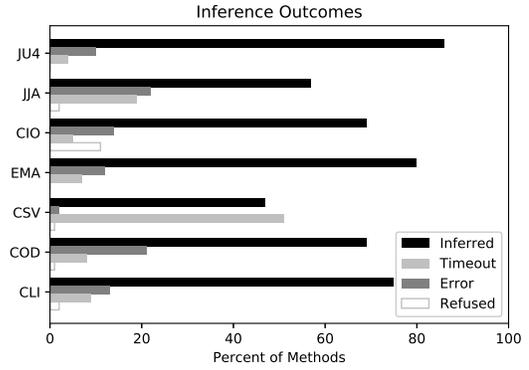}
   
  \caption{A summary of inference outcomes for our technique on our 7 test libraries. }
  \label{fig:inference-breakdown}
\end{figure}

For each of the libraries we categorized the status of an inference attempt in one of four different categories. We give an explanation of the categories in this section as well as provide some analysis of our findings; see Figure~\ref{fig:inference-breakdown}.

The status \textit{Inferred} indicates that our technique successfully inferred a specification for a method. We define success as containing at least one postcondition in the form of an \texttt{ensures} or \texttt{assignable} clause (in the case a frame axiom is necessary). Other clauses may be (and often are) present. For all experiments our technique succeeded in producing a specification more than $74.0\%$ of the time (overall), and in its worst performance produced a specification $48.1\%$ (in Commons-CSV). Our technique's best success in producing specifications was on JUnit4 (92.5\%); this was largely due to JUnit4 having significantly smaller control flow graphs compared to other libraries. In our analysis we found that the CFGs for JUnit4 were less than 400 nodes, except for one case of 594 nodes.

\begin{figure}[t]
  \centering
\includegraphics[width=.8\linewidth]{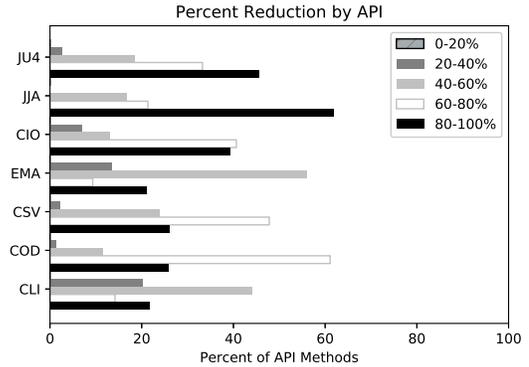}   
\caption{A summary of specification length reduction for our technique on our 7 test libraries. }
\label{fig:reduction-breakdown}
\end{figure}

A status of \textit{Timeout} indicates that the inference process was aborted before inference could complete. For this paper we used a timeout of 300 seconds (5 minutes). We determined this timeout through repeated experimentation with different timeouts ranging up to 20 minutes. In our initial tests we determined that inference attempts that did not complete within 5 minutes were not able to complete in 20 minutes either. While higher settings for timeout timeouts might reduce the number of specifications that fail to be inferred, the combined timeout was only about $9\%$. When a timeout occurs the intermediate results are discarded and not included in our remaining analyses. In the results reported in Figure~ \ref{fig:inference-breakdown} we can see that the two worst performing subjects in terms of timeout are Commons-CSV ($51.3\%$) and JSON-Java ($23.0\%$). Unlike the other test candidates, both Commons-CSV and JSON-Java are both parsers that contain deeply nested code. In all other libraries timeout performance was excellent and perhaps these results suggest that inferring specifications for parsers, since they are inherently very recursive, is more difficult than inferring specifications for general purpose code.

A status of \textit{Refused} means that our technique did not attempt to infer a specification, because the control flow graph (CFG) size was larger than a preset limit (a CFG size of 500 nodes). Similar to the timeout parameter, larger CFGs typically take much longer to infer. In our evaluation a size of 500 proved to be a reasonable choice, since the number of refused methods represented only $3.9\%$ overall. This number is partially inflated by the unusually high number of refused methods in Commons-IO. This is explained by higher CFG complexity relative to the other libraries in the test suite. This additional complexity comes from the way the verification conditions for exceptional code are generated. Although we are not inferring the specifications for the exceptional specifications cases, the exceptional information exists in the CFG that our technique analyzes to infer the normal specification cases. Since Commons-IO deals with input/output related functions it has an unusually high number of exceptions. This greatly inflates the size of the CFG, which made fewer of its methods usable for our study. This effect could be mitigated if the exceptional nodes were removed from the CFG prior to inference but such a change would make it then impossible to later infer the exceptional behavior of methods. 


A status of \textit{Error} means that our technique encountered an internal error during inference. We manually investigated these errors and found them to be generated from current limitations in OpenJML itself. For example, certain features, such as enumerated types, are not currently supported in OpenJML (although they are valid in JML itself). Our implementation is based on OpenJML 0.8.12; we expect to be able to reduce the amount of internal error our tool encounters as errors are corrected in OpenJML.

\subsection{Efficiency of Reduction}
\label{sec:efficiency-reduction}

\begin{figure}
 
  \centering
\includegraphics[width=\columnwidth]{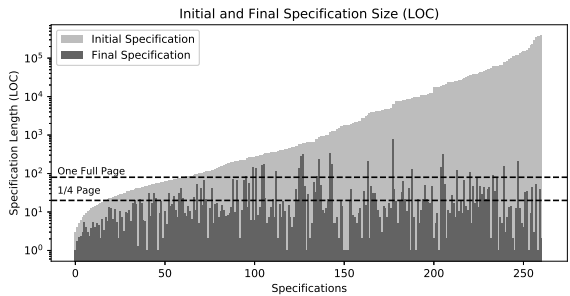}

\caption{The initial and final specification length for all inferred specifications (all test libraries combined, binned to 262 equal width bins).  Light lines indicate length before applying our technique; dark lines show our technique's effect on length. In this paper we consider 1/4 page to be equal to 20 lines.}
\label{fig:red-blue}
\end{figure}


As discussed in Sections~ \ref{sec:introduction} and \ref{sec:technical-approach}, one of the problems impacting specification inference by symbolic execution (and therefore predicate transformers) is the length of the resulting inferred specification. In this section we will evaluate how effective our technique was at reducing the length of specifications inferred by symbolic execution.

In Figure~ \ref{fig:red-blue}, we show the length of the initial and final inferred specifications in terms of lines of specification. Additionally, since short specification length can be a desirable quality from a software engineering and readability perspective, we also provide two reference lines: one at one full page (80 lines) and one at a quarter page (20 lines). In our analysis, $95.0\%$ of specifications fell below one full page length and $84.6\%$ fell below one quarter page length. This suggests that our techniques for reducing specification size were effective in reducing the size of most of the inferred specifications.

The results in Figure \ref{fig:red-blue} give an overview of all inferred specifications. However, our technique's effectiveness at reducing the final specification length of individual code bases varied significantly. In Figure \ref{fig:reduction-breakdown}, we can see a breakdown of our technique's effectiveness at reducing the final size of inferred specifications. Of the most interest are 80-100\% category (the \textit{most} reduction), and the 20-40\% reduction category (the \textit{least} reduction). Although we created a category for it, none of the code bases contained specifications that were reduced in the 0-20\% category. In our experiment, we found that the 80-100\% category contained 41.6\% of all methods. In Figure \ref{fig:reduction-breakdown} we see that the worst performer (least reduction) was Commons-Email. We did further analysis of this phenomenon where we looked at the length of the methods being inferred, the control flow depth, and the resulting reduction classification. We did not manage to find a relationship between these variables, which may suggest that these contracts failed to reduce more significantly because they described essential (non-reducible) logical states rather than non-practical artifacts as described in Section \ref{sec:technical-approach}.

\subsection{Complexity of Inferred Specifications}
\label{sec:compl-inferr-spec}

Criticisms of prior work on specification inference have included that the inferred specifications were either too short and not descriptive enough or template based and not expressive enough to capture the meaning of programs not falling within the parameters of the templates. We examine complexity in three different ways. First, in Figure \ref{fig:red-blue} we give metrics on the variation in the length of specifications produced by our technique. To quantify the distribution of types of clauses our technique was able to infer, in Table \ref{table:complexity}, we report on the variation of clauses present in the specifications inferred by our technique for each of our target codebases. In Table \ref{table:complexity} we can see that inferred specifications were comprised of mostly \texttt{requires} clauses, followed by roughly 4 times less \texttt{ensures} clauses as well as smaller numbers of \texttt{pure} and \texttt{assignable} clauses. Our technique is able to handle quantifiers such as \texttt{forall}, however we did not examine its distribution in the inferred specifications.  In Figure \ref{fig:far-nesting}, we detail the specification case nesting present in the inferred specifications. This is discussed further in Section \ref{sec:far-performance}.

\renewcommand{\arraystretch}{1.1}
\begin{table}[t]
	\caption{Summary of JML clauses in inferred specifications.}
	\label{table:complexity}
  \begin{tabularx}{\linewidth}{Xrrrrrr}
    \toprule  
    API &  Methods & \texttt{requires} & \texttt{ensures} & \texttt{assignable}  & \texttt{pure} \\ \midrule
    JU4         & 1,230  & 845   & 631 & 153  & 588 \\
    JJA     & 200   & 209   & 183 & 24   & 40  \\
    CSV   & 158   & 70    & 91  & 34   & 23  \\
    CLI   & 194   & 1,301  & 422 & 118  & 27  \\
    COD & 509   & 1,533  & 569 & 105  & 90  \\
    EMA & 192   & 7,139  & 622 & 191  & 47  \\
    CIO    & 955   & 1,483  & 681 & 270  & 641 \\
    \midrule 
    Total & \textbf{2,331} & \textbf{12,580} & \textbf{3,199} &  \textbf{895} & \textbf{1,456}   \\
    \midrule
    Ratio & - & 5.39 & 1.37 & 0.38  & 0.62 \\

    \bottomrule 
  \end{tabularx}
\end{table}

\subsection{Performance of Inference}
\label{sec:perf-infer-proc}

To better understand the runtime performance characteristics of our technique we collected telemetry data about its performance. In our analysis, we found that the data was tightly clustered  between $10^1$ and $10^2$ ms. Note that inference time in our experiment was limited to 5 minutes ($3 \times 10^5$ ms). Many of the observed data points fell within this region, suggesting a linear fit with the exception of some very large control flow graphs which fell in the region bordering the timeout. These data points were identified as belonging mostly to Commons-IO and due largely to the number of exceptional flows present in the resulting control flow graphs (see Section~ \ref{sec:effect-infer} for more details). As discussed in Section~ \ref{sec:effect-infer}, using the timeout threshold of 5 minutes with the current performance characteristics allowed us to infer more than 75\% of the methods we considered for this study with an average inference time of just 2 seconds per method.

\subsection{Performance of our technique}
\label{sec:far-performance}

\begin{figure}[h!]
    \centering
    \includegraphics[width=0.7\linewidth]{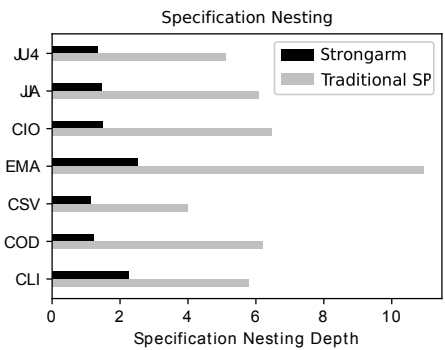} 

  

  \caption{The effect of our technique on the nesting of inferred specifications }
  \label{fig:far-nesting}
  
\end{figure}

To better understand the performance of our technique with respect to more standard techniques such as removing tautologies, we conducted further evaluation to study our technique. We conducted our study by taking the same 7 libraries used throughout this section and examining the effect of running all of the standard analysis types (Section \ref{subsec:implementation-details})  with the exception of our technique in one run and all of the analysis types \textit{with} our technique added back into the analysis pipeline. To understand the characteristics of specifications before and after our technique is applied to them we observed 4 different metrics. The first metric we observed was \textbf{specification nesting}. Specification nesting is defined as the lexical depth of a specification. In Figure \ref{fig:far-nesting}, we present the performance of our technique in reducing nesting of specifications. In this case of all libraries, our technique achieved a reduction in nesting, producing an overall average reduction of 73.8\%.

In our examination of the \textbf{percent reduction} aspects of our technique, we found our technique's effect on specifications is not necessarily to produce large reductions in the length (as measured in lines) of specifications; our technique's contribution to the length (in lines) is small compared to the contribution of the other steps (our technique contributes an additional $\approx 10\%$ reduction overall).

\begin{figure}[t]

  \begin{subfigure}{.5\linewidth}
    \includegraphics[width=\linewidth]{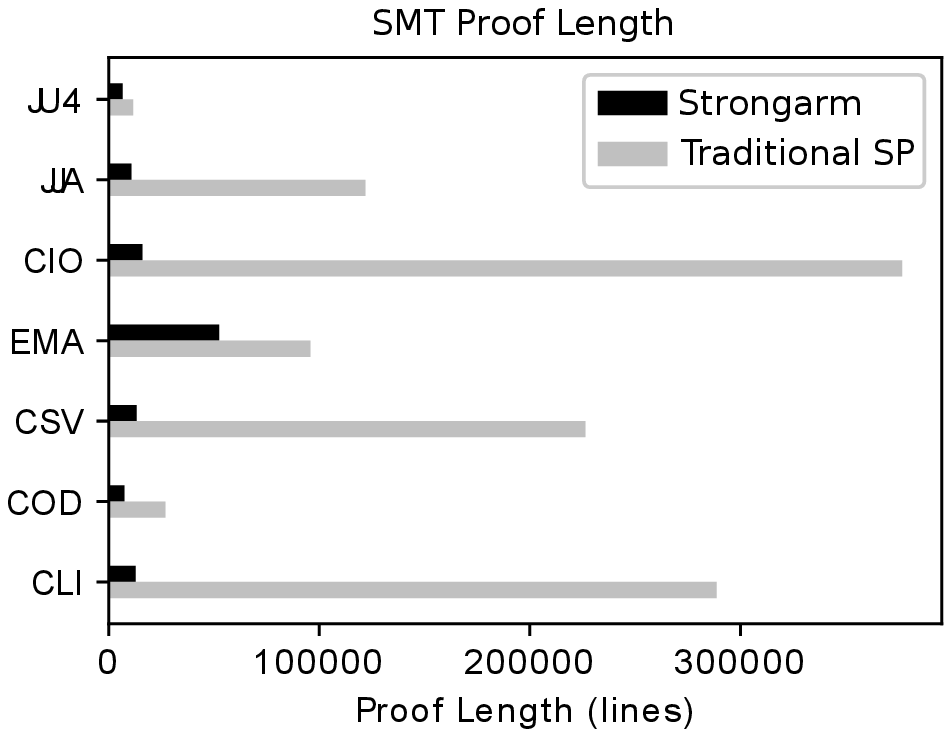} 
  \caption{} 
  \label{fig:far-proof-length}
  \end{subfigure}%
  \begin{subfigure}{.5\linewidth}
    \includegraphics[width=\linewidth]{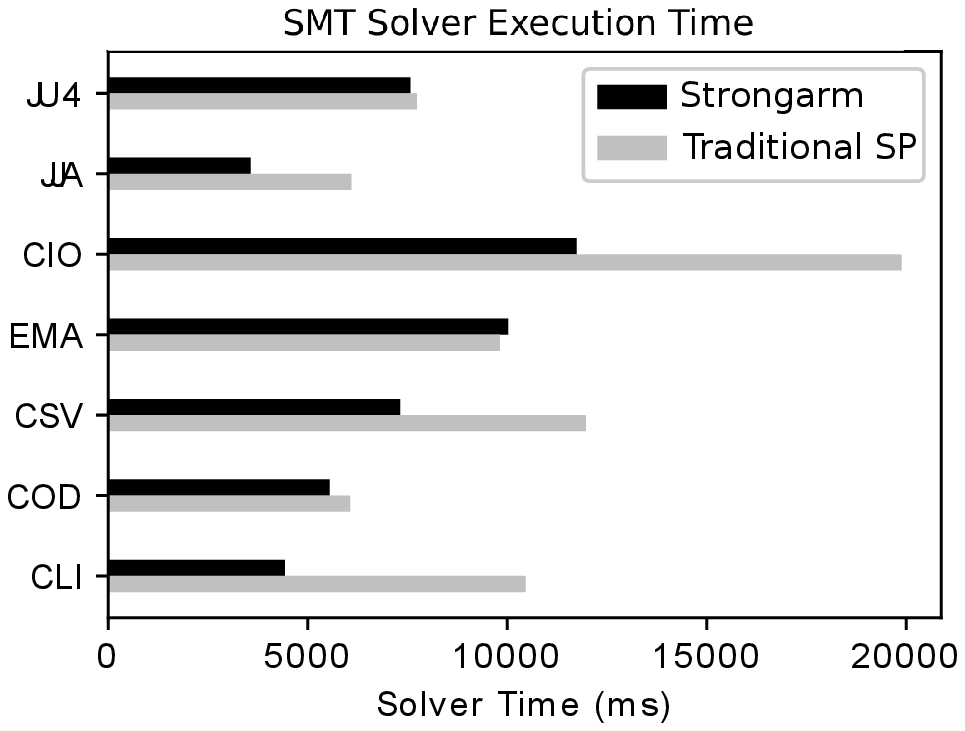}
  \caption{}  
  \label{fig:far-proof-time}
  \end{subfigure}%
    
  \caption{The effect our technique on (a) the length of the SMT conditions generated from specifications (b) the time needed to check the inferred specifications.}
\end{figure}

Ultimately, for a specification to be most useful, it should be used to verify the implementation of the code it specifies. In Figure \ref{fig:far-proof-length}, we examine the impact of our technique on the \textbf{proof length}, i.e., the length of the SMT conditions generated that are needed to verify the correctness of a specification in relation to its implementation. In Figure \ref{fig:far-proof-length} we can see that our technique has a large impact on the size of the generated SMT proofs for the code samples we studied. Overall, \textit{our technique reduces the size of the generated SMT conditions by 76.7\%}.

In the previous paragraph we saw that our technique is effective for reducing the size of the proofs required to verify implementations of inferred specifications. To determine the impact of the reduced length on the time to run the SMT solver on these proofs, we studied the \textbf{solver times} with and without our technique. In Figure \ref{fig:far-proof-time} we show the time taken to prove the specifications inferred with and without our technique. As can be seen in all cases, our technique reduces the time taken to verify the SMT conditions. Across all libraries, our technique reduces the prover execution time by  26.7\%.


\subsection{A Study of the Inferred Specifications} \label{ln:practical}

To determine if the specifications produced by our technique were useful to human readers, we conducted a small study to examine their usefulness. In general, we believe that practical specifications are more useful for human readers than those that are not, since they tend to be shorter and more to the point. However, we defined additional criteria against which we designed our study. 

Namely, we selected the following criteria and say that a specification is \textit{practical} if they:

\begin{itemize}
\item
do not contain redundant formulas, 
\item  
do not contain unsatisfiable formulas,
\item  
do not contain tautological formulas,
\item  
specify frame axioms \cite{Borgida:1993:LNE:257572.257636}), 
\item  
specify when a method is pure (has no side-effects), 
and
\item  
only use names that are visible to a method's clients. 
\end{itemize}


\textbf{Methodology.} Using the specifications inferred by our technique, we conducted a survey of people familiar with the Java Modeling Language (JML)~\cite{leavens_preliminary_2006}. The survey used 12 pairs of method specifications inferred by our technique; to highlight the aspects of our definition of practical specifications, one element of each pair had one aspect, e.g., removal or unsatisfiable formulas, disabled. 
The survey was completed by 25 people, 18 of which said that they ``definitely'' had experience reading JML and 7 of which had some experience with JML.  All correctly answered a simple question about JML that tested their understanding of a method specification with two specification cases.
(A method \emph{specification case} in JML is a pre- and postcondition specification, with optional frame axioms, that must be satisfied whenever the case's precondition is true when the method is called.)
The questions in the survey were all simplifications of pairs of output given by our tool, one of which was not processed to remove a single impractical feature (according to the above definition).

\textbf{Results.} An unsatisfiable precondition, such as \lstinline/!true/, causes the specification case it appears in to be useless.
87.5\% of the survey respondents preferred a specification without unsatisfiable preconditions, including preconditions that required non-null fields to be null.
Another example of unsatisfiable specifications comes when a specification case has two mutually-contradictory clauses; in this case all 
respondents preferred a specification without such unsatisfiable combinations of clauses (75\% of them strongly so).

According to the survey, 95\% of the respondents preferred a specification without the tautological postcondition \lstinline!true == true!.
Also, 62.5\% of the respondents preferred a specification without tautologies such as \lstinline!requires true! and more subtle tautologies involving non-null declarations. Another 87.5\% preferred a specification without subtle redundancies such as \lstinline/"No resource defined" != null/.
In another question, 80\% of the respondents preferred a specification without redundant clauses requiring non-null fields to be not null.
In another question, 93.3\% of the respondents preferred a specification without duplicated specification cases.

Regarding frame axioms, the survey contrasted a specification
without the frame \lstinline!assignable this.name!
with a nearly identical specification with the assignable clause.
Although the specification without the frame axiom is shorter, 86.7\% of the respondents preferred the specification with the frame axiom (53\% strongly so).

In JML a pure method is one without any side-effects, and is specified by the \lstinline!pure! keyword. The \lstinline!pure! keyword also functions as a strong frame axiom.
66.7\% of the respondents preferred a specification to one just like it but without the keyword \lstinline!pure!.


In sum, the majority of the survey respondents preferred specifications that satisfy our notion of a ``practical'' specification (the kind our technique produces), even if the differences were rather subtle.


\section{Related Work}
\label{sec:related-work}

In addition to the work discussed in Section \ref{sec:introduction}, there are several other systems and papers relevant to this paper. The closest related work to Strongarm is the Houdini system by Flanagan and Leino \cite{flanagan_houdini_2001}. Houdini targets ESC/Java with the goal of statically inferring specifications for Java. Similarly, in our work we target OpenJML, a successor to ESC/Java. However, rather than symbolically computing specifications as Strongarm does, Houdini instead applies templates, i.e., commonly-found specification patterns; to search for a specification, Houdini tries all of these patterns and checks the candidate specifications using ESC/Java. In contrast to Strongarm, Houdini also attempts to infer object invariants using the same mechanism, which Strongarm does not yet attempt. 

Flanagan and Saxe's work on compacting verification conditions \cite{flanagan_avoiding_2001}, unlike Strongarm, only deals with two types of duplication: multiple assignment and propagation of accumulated formulas.  Strongarm can reduce these problems globally in the specification whereas the approach of Flanagan and Saxe is restricted to the branch level. Furthermore, Strongarm uses a pluggable equivalence relation ($\sim$, see Section \ref{subsec:flattening}), which generalizes their work.
 
In contrast with Strongarm (and Houdini), Daikon \cite{ernst_daikon_2007} is a runtime approach. To produce specifications, Daikon requires a test suite that calls the code that is being targeted for inference. Even when such a test suite is available, dynamic inference can fail, since code coverage and branch coverage are typically not sufficient to produce specifications \cite{harder_improving_2003,gupta_new_2003}, as many test suites focus on corner cases and are thus not suitable for specification inference \cite{nimmer_invariant_2002}.

Similarly, in their work on discovering relational specifications, Smith et al. discover specifications by looking at program outputs \cite{smith_discovering_2017}. Our approach is fully static does not require running the code. 
   
In their work on API usage error detection, Murali et al. investigate detecting API usage errors  using Bayesian inference \cite{murali_bayesian_2017}. Our approach does not use machine learning to generate its specifications and therefore does not require training examples to infer specifications.

Nguyen et al.~\cite{nguyen_mining_2014-1} infer method preconditions, by examining call sites. Our approach is different in that Strongarm computes postconditions based on each method's code, not preconditions based on calling code.

\section{Conclusion and Future Work}
\label{sec:conclusion}

Our approach to inferring practical postconditions takes advantage of structure the logical formulas that result from the SP predicate transformer. We evaluated our technique on 7 popular Java libraries and found that 95.0\% of the inferred specifications were less than 1 page long and 84.6\% were less 1/4 of a page.  
Such concise specifications have the potential aid many areas of software engineering. 
Our future work will evaluate these implications.


\balance

\bibliographystyle{plain} 
\bibliography{main} 



\end{document}